# Deep-learning-based Radiomics on Mitigating Post-treatment Obesity for Pediatric Craniopharyngioma Patients after Surgery and Proton Therapy

Wenjun Yang[1], Chia-Ho Hua[2], Tina Davis[2], Jinsoo Uh[2], Thomas E. Merchant[2]

Department of Radiation Oncology,

[1]University of Iowa Health and Clinics

200 Hawkins Dr.

Iowa city, IA 52242

[2]St. Jude Children's Research Hospital

262 Danny Thomas Pl.

Memphis, TN 38105

**Conflict of Interest Notification**

There are no conflicts of interest in this study.

**Abstract**

**Background:** Craniopharyngioma is a benign brain tumor often occurs in children and adolescents. Limited surgery following by irradiation effectively controls the treatment outcome reaching over 80% survival rates for 10 years. However, the quality of life for the pediatric patients lacks quantitative management, although clinical factors including age, sex, race, shunt placement, and growth hormone have been recognized key features correlated with post-treatment obesity.

**Purpose:** Mitigating post-treatment obesity for pediatric craniopharyngioma patients will improve their quality of life with a long span expectation. We developed an artificial neural network (ANN) combining radiomics with clinical and dosimetric features to predict the extent of body mass index (BMI) increase after surgery and proton therapy, with advantage of improved accuracy and integrated key feature selection.

**Methods and Materials:** Uniform treatment protocol composing of limited surgery and proton radiotherapy was given to 84 pediatric craniopharyngioma patients (aged 1-20 years). Post-treatment obesity was classified into 3 groups (<10%, 10-20%, and >20%) based on the normalized BMI increase during a 5-year follow-up. We developed a densely connected 4-layer ANN with radiomics calculated from pre-surgery MRI (T1w, T2w, and FLAIR), combining clinical and dosimetric features as input. Accuracy, area under operative curve (AUC), and confusion matrices were compared with random forest (RF) models in a 5-fold cross-validation, with robustness tested by adding random radiomic noises. The Group lasso regularization in value function optimized a sparse connection to input neurons, and thus identifying key features for the obesity prediction.

**Results:** Classification accuracy of the ANN reached above 0.9 for T1w, T2w, and FLAIR MRI. Confusion matrices showed high true positive rates of above 0.9 while the false positive rates were

abstract
below 0.2. Approximately 10 key features including radiomics of intensity minimum, kurtosis, skewness; clinical features of age, sex, race; and dosimetric features of target and hypothalamic doses were selected for T1w, T2w, and FLAIR MRI, respectively. The ANN improved classification accuracy by 10% or 5% when compared to RF models without or with radiomic features. The classification accuracy was robust to noise while the key features collection were relatively prone to the radiomic noise with half magnitude of connection weights.

**Conclusion:** The ANN model improved classification accuracy on post-treatment obesity compared to conventional statistics models. The clinical features selected by Group lasso regularization confirmed our practical observation, while the additional radiomic and dosimetric features could serve as imaging markers and mitigation methods on post-treatment obesity for pediatric craniopharyngioma patients.

**Introduction**

Craniopharyngioma is a benign tumor with locally aggressive features that often arises in the suprasellar region [1]. The invasive compression in the hypothalamic-pituitary axis [2, 3] results in growth hormone deficiency and occasionally diabetes insipidus [4, 5]. Gross total resection induced hypothalamic injury [6] could aggravate hormone deficiency [7] and hence enhance the risk of post-treatment obesity [8, 9]. Thus, limited surgery with complementary irradiation to the tumor bed is currently the main-stream treatment for pediatric craniopharyngioma [8]. We performed a treatment protocol with limited surgery and proton therapy for a pediatric patients cohort who have long life-expectation: the 10 years event free and overall survival rates reached 80% and 90%, respectively for limited surgery plus radiation and radiation treatment alone [3]. Based on the promising treatment outcome, we developed deep learning models in this study to monitor and mitigate the post-treatment side effects to further improve the quality of life for the pediatric patients.

To monitor post-treatment obesity and cognitive function decline as the major side effects, we followed up 5 years after treatment recording their cognitive score, growth hormone level, body mass index (BMI), etc. Complementary growth hormone therapy was prescribed to specific patients based on the follow-up tests. In this study, we focused on predicting post-treatment obesity and mitigation methods. Although direct injury to hypothalamus by occurrence of tumor and the following surgery and proton therapy may be obvious contributing factors [9], the mechanism of intrinsic resistance to radiation therapy among patients who maintained steady BMI remained unclear [10]. We observed that clinical factors including patient age, sex, shunt placement, race, and growth hormone therapy were highly correlated to the occurrence of post-treatment obesity [3], while there lack quantitative modelling to the best of our knowledge.

Radiomics is a statistical model with input of high dimensional features calculated from macro- or microscopic medical imaging properties to predict clinical endpoints [11]. The macroscopic features including target shape, volume, dimensions, etc. are commonly used in diagnostic imaging, while the microscopic texture features including energy, entropy, kurtosis, skewness, etc. boost the model's prediction by adding complementary information [12-15]. The biological mechanism of radiomics is often the correlation between genetic expression [16] and microscopic imaging texture, while this topic is beyond the scope of this study. Despite improved prediction accuracy, the high-dimensional input of radiomics often results in overfitting and lack of interpretation for the decision strategy. We developed a densely connected artificial neural network (ANN) with compound input features composing of radiomics, clinical, and dosimetric factors to classify the extent of BMI increase during the 5-year follow-up. This is a pilot study on post irradiation obesity for pediatric patients with craniopharyngioma, and we further innovated the ANN by integrating a Group lasso regularization in the value function to identify key features among the high-dimensional input space as imaging markers and mitigation methods. The performance of ANN was compared with random forest (RF) models and the robustness was justified by adding varying magnitude of radiomic noise.

**Methods and Materials**

**Patients**

Treatment protocol composing of limited surgery and proton therapy was performed on a patient cohort of 84 pediatric patients (44 females, 40 males, aged 1-20 years), with IRB approval by our hospital. Cerebral shunt was placed for patients with high cranio-spinal fluid pressure for limited surgery, followed by conformal proton therapy on the tumor bed. The clinical target volume (CTV) was defined by a 0.5-cm margin surrounding the postoperative tumor bed, and planning target volume (PTV) was expanded by a 0.3-cm margin surrounding the CTV. Two oblique conformal proton beams delivered 54 Gy (RBE) in 30 fractions. We followed up with the patients every 3 months during the first year after proton therapy recording clinical (BMI), laboratory (growth hormone level), and imaging (MRI) evaluation, and then every 6 months through approximately year 5 (median = 5 years, range = 1-5 years).

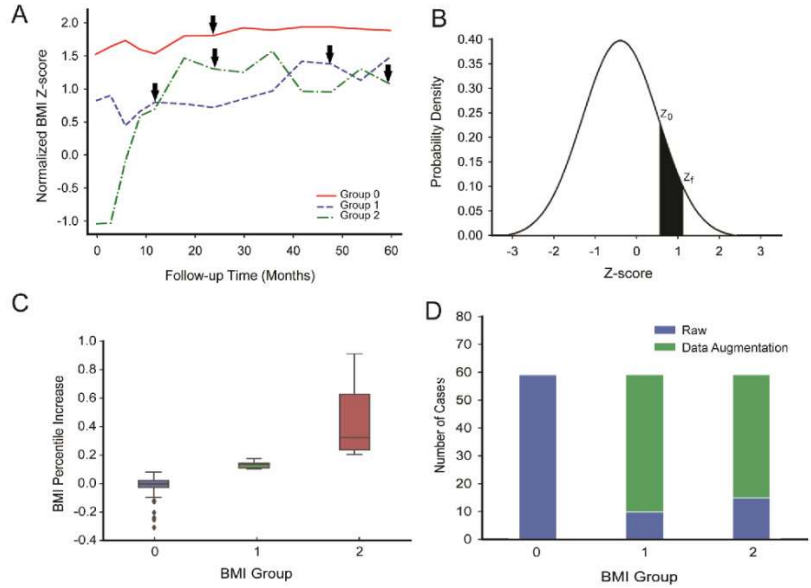

Fig. 1. Post treatment obesity definition and classification methods. (A) Three representative BMI growth charts for the 3 groups over a 5-year follow-up period. The curve for group 0 (red) was essentially stable; the curve for group 1 (blue) increased by BMI z-score of 10-20%, and the curve for group 2 (green) grew by more than 20%. The black arrows denote the starting and stopping dates of growth hormone replacement therapy, while the single arrow for group 0 indicated a one-week therapy. (B) Definition of BMI z-score increase: the swiped area between $Z_0$ and $Z_f$ indicates the normalized BMI percentile increase between proton therapy and the final follow up measurement. (C) Box plot of the BMI percentile increases for group 0-2. The center bar in box plot indicates the median, while the lower and upper boundaries indicate the 25% and 75% percentile. Patients in group 0-2 had BMI increase <10%, 10-20%, and >20%, respectively. (D) SMOTE data augmentation for groups 1 and 2: a linear interpolation was performed in groups 1 and 2 to balance the number of cases in all 3 groups, with a blue bar indicating original cases and green bar indicating synthetic ones.

Patients were evaluated for growth hormone deficiency and adrenal insufficiency via provocative testing before undergoing proton radiotherapy. Other hormone axes were evaluated before and after treatment by serum screening tests and clinical evaluations. Hormone replacement therapy was initiated when clinically indicated as marked by the black arrow in Fig. 1(A).

**Endpoints**

We defined the post-treatment obesity as the relative shift of normalized BMI z-score adjusted for age and sex in accordance with the growth chart produced by the Centers for Disease Control and Prevention (CDC) [17]. The relative shift was depicted as the shaded area between initial and follow up BMI z-score in the Gaussian distribution as shown in Fig. 1(B). The normalized BMI z-score was calculated from raw value adjusted with age- and sex-dependent parameters as follows:

$$Z = \frac{\left(\frac{X}{M}\right)^L - 1}{LS}, L \neq 0,$$

$$Z = \frac{\ln\left(\frac{X}{M}\right)}{S}, L = 0 \quad (1)$$

where $Z$ is the normalized BMI z-score, $X$ is the measured BMI raw data, $L$ is the power in the Box–Cox transformation, $M$ is the median value for a specific age and sex group, and $S$ is the generalized coefficient of variation. Parameters $L$, $M$, and $S$ are dependent on age and sex as listed in the CDC growth chart, which covers an age range from 2 to 20 years [17]. For 4 patients whose ages were outside this range (0.88, 1.74, 1.98, and 20.14 years), the $L$, $M$, and $S$ parameters for ages 2 and 20 years were applied as appropriate.

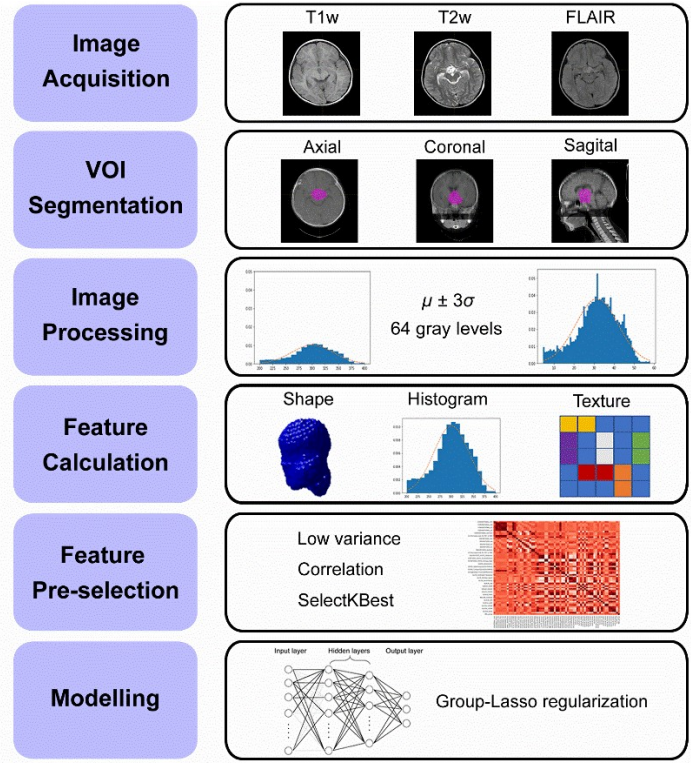

Fig. 2: Data processing pipeline. The gross tumor volume was segmented on pre-treatment T1w, T2w, and FLAIR MRI by a radiation oncologist in the treatment planning software. Image preprocessing was implemented to normalize the multi-centric acquisition. Different levels of features were calculated within the segmented tumor volume. A 4-layer densely connected artificial neural network with Group lasso and dropout regularization was trained for prediction and feature selection.

We classified BMI increase into 3 groups based on the area between $z_0$ and $z_f$ as shown in Fig. 1(B-C). Group 0 patients had a BMI z-score increase of less than 10 percentile between $z_0$ and $z_f$; group 1 patients had shaded area of between 10 and 20 percentile; and group 2 patients had more than 20 percentile shifts. The classification groups were labeled as the output of the ANN model. The box plot in Fig. 1(C) illustrates the range and frequency of BMI z-score increase in the 3 groups, respectively, while the trending curves during the 5-year follow-up were illustrated in Fig. 1(A) with 3 examples for the corresponding group.

**Image Processing and Feature Extraction**

Preoperative MRI for the pediatric craniopharyngioma patients were scanned in a variety of imaging centers as summarized in Table 1. Image features were calculated from 72 T1-weighted (T1w), 79 T2-weighted (T2w), and 67 FLAIR MRI based on availability among the 84 patients in cohort. Radiation oncologists manually delineated the preoperative gross tumor volume (GTV) on MRI. Rigid registration between the planning CT and MRI, image preprocessing, and feature calculation were performed using the LIFEx software (Institut Curie Research Center, Orsay, France) [18], as stepped in Fig. 2. The preprocessing [19] to unify the multi-imaging centers was performed by a resampling of image resolution to 1 mm × 1 mm × 3 mm [20, 21], and a resampling of gray levels to 64, centered and windowed by the mean $\pm 3 \times$ the standard deviation of the segmented GTV [22].

Macroscopic or first order radiomic features were calculated with GTV shape, intensity histogram, conventional index features, etc. Microscopic or second-order textures included the gray-level co-occurrence matrix (GLCM) [23], gray-level run-length matrix (GLRM) [24], neighborhood gray-level different matrix (NGLDM) [25], and gray-level zone-length matrix (GLZLM) [26], etc. A total number of 105 radiomic features were calculated in 3 dimensions within the GTV. Highly correlated clinical features based on our practice were listed in Table 1 and the statistics of biological equivalent dose to hypothalamus and target was input as dosimetric features. Separate models were built for T1w, T2w, and FLAIR MRI, respectively, due to the limited availability of MRI sequences to each patient.

**Table 1. MRI scanners and clinical relevant features of the patient cohort**

| MRI scanner | 1.5 T | 3 T | Age | | Sex | | Race | | Procesdure | | Growth hormone (mg) | |
|---|---|---|---|---|---|---|---|---|---|---|---|---|
| | | | 0-5 | 12 | Male | 40 | Black | 15 | Shunt placement | 35 | 0-5 | 41 |
| Siemens | 33 | 11 | 5--10 | 37 | Female | 44 | White | 56 | Endoscopic resection | 49 | 5--10 | 14 |
| GE | 19 | 7 | 10--15 | 21 | | | Asian | 2 | | | 10--15 | 10 |
| Philips | 11 | 3 | 15--20 | 14 | | | Pacific Islander | 1 | | | >15 | 19 |
| | | | | | | | Other | 10 | | | | |

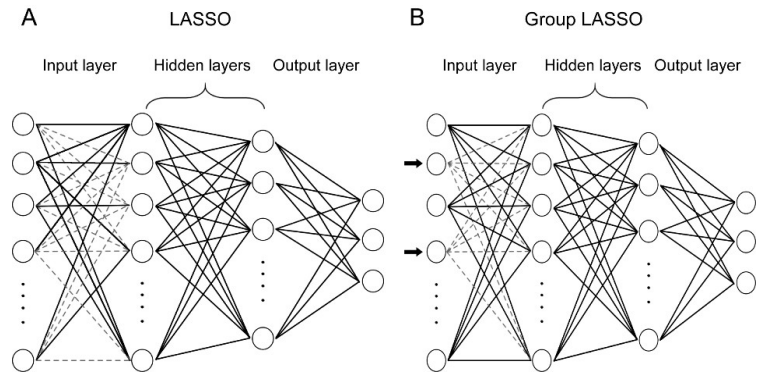

Fig. 3: Artificial neural network regularized by Group lasso penalty. Conventional Lasso (A) randomly regulates connections between neurons, whereas Group lasso (B) shrinks connections to certain neurons as a bundle, as indicated by the arrows.

**Modeling**

Radiomic, clinical, and dosimetric features were filtered by a variance threshold to drop the low-informational features. Sixty-two radiomic, 5 clinical, and 13 dosimetric features were input to a 4-layer densely connected ANN model, with 80, 40, 20, and 3 neurons in each layer. The first and last layer contained the number of neurons that equals to input features, and BMI increase groups, respectively, as described in the endpoints section. The hidden layers contained a decreasing number of neurons for information extraction. The Group lasso regularization integrated in the value function identified key features as shown in Fig. 3: conventional lasso regularization randomly shrunk the weights of connection between each layer as shown in Fig. 3(A), while the Group lasso regularization minimized the weights of connections to individual neurons as a bundle as shown in Fig. 3(B), such that the key features were identified by ranking the sum of weights for each neuron in the input layer: final features were selected by a 2-sample *t*-test (*P* < .005) with a testing hypothesis that the weight (L2 norm of the column vector in the connection matrix between the input and following hidden layer) of each feature was higher than the average weight of all features over a 5-fold cross-validation. The Group lasso regularization [27, 28] was formulated as follows:

$$R_{gl}(w) = \sum_{w \in G} \sqrt{D} ||w||_2, \qquad (2)$$

where *w* is the column vector in weight matrix, *G* is the total set of weights in the input and hidden layers, and *D* is the dimensionality of vector *w*. A 5-fold cross-validation was used to evaluate the ANN model in classification accuracy, confusion matrix, and area under the receiver operating characteristic curve (AUC). The synthetic minority oversampling technique (SMOTE) [29] was applied to augment the number of unbalanced samples in groups 0 and 1 as shown in Fig. 1(D) due to the limited number of true positives in the patient cohort.

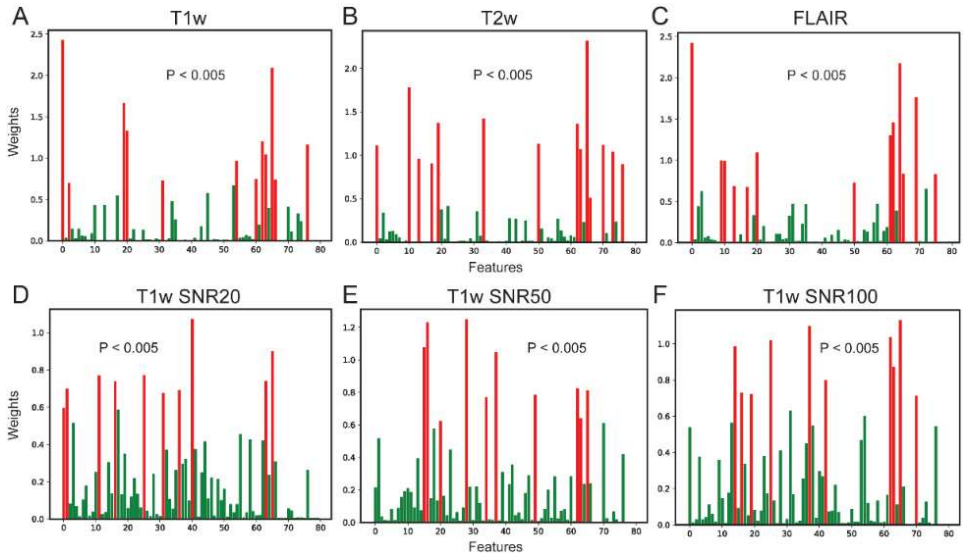

Fig. 4: The weights of the features selected by the Group lasso regularization. A sparse distribution was regularized for the T1w, T2w, and FLAIR MRI as shown in (A-C), while the noise with a variety of SNR compromised the sparsity by half the magnitude as shown in (D-F). The final features are selected with a *t*-test (*P* < .005) as indicated by the red bars.

We compared the classification accuracy of ANN and RF models with and without radiomic features. The enhanced classification accuracy for RF model with radiomic features demonstrated the effectiveness of radiomics in addition to conventional clinical and dosimetric features, while the comparison between ANN and RF models with the same set of all features justified the superior performance of ANN over RF. We further tested the robustness of ANN by adding radiomic noise with SNR 20, 50, and 100, respectively.

**Results**

The average 5-fold accuracy of classification for T1w, T2w, and FLAIR MRI with ANN achieved 0.90, 0.93, and 0.92, respectively. For RF models with clinical and dosimetric features only, the accuracy decreased to 0.79, 0.81, and 0.77 for T1w, T2w, and FLAIR MRI. By adding radiomics features to the RF models, the accuracy was improved to 0.85, 0.85, and 0.91 for corresponding MRI. The comparison of classification accuracy for all models was summarized in Table 2. Average AUC of the 5-fold test for ANN achieved 0.98, 0.99, and 0.97 for T1w, T2w, and FLAIR MRI, respectively.

**Table 2. Classification accuracy of ANN and RF models with different set of input features**

| MRI | RF (Clinical + Dosimetric) | RF (All features) | ANN (All features) | ANN SNR 100 | ANN SNR 50 | ANN SNR 20 |
|---|---|---|---|---|---|---|
| T1w | 0.79 | 0.85 | **0.9** | 0.88 | 0.87 | 0.86 |
| T2w | 0.81 | 0.85 | **0.93** | **0.93** | **0.92** | 0.88 |
| FLAIR | 0.77 | 0.91 | **0.92** | **0.92** | **0.93** | 0.88 |

Key features selected by the Group lasso regularization of the corresponding MRI sequences were summarized in Table 3, while the relative weight of each feature was illustrated in Fig. 4(A-C). The selected key clinical features eg. age, sex, race, shunt placement, and growth hormone therapy confirmed our clinical observation. Several dosimetric features including target minimum dose, hypothalamus minimum and median dose were key features for certain MRI. Key radiomics features eg., minimum gray level, histogram skewness, and kurtosis were commonly selected for all MRI.

Confusion matrices in Fig. 5 (A-C) demonstrated the classification results for individual group, where the principal component was the true positive while the other elements were misclassification rates to other groups. For T1w, T2w, and FLAIR MRI, the true positive rate reached around 0.8 for group 0, while achieved above 0.95 for group 1 and 2. These results suggest a clinically favorable higher true positive (0.95) than true negative rates (0.8) for post-treatment obesity prediction.

**Table 3. Key features selected with the Group lasso regularization in ANN**

| MRI | Key features selected by Group lasso regularization |
|---|---|
| T1w | min, Skewness, ExcessKurtosis, HISTO_Skewness, PARAMS_BinSize, GLCM_Contrast(Variance), GLZLM_LZE, GLZLM_LGZE, Race, Age, Shunt, Hypo_dose_min |
| T2w | min, Skewness, HISTO_Skewness, HISTO_Entropy_log10, HISTO_Entropy_log2, GLCM_Energy(AngularSecondMoment), GLRLM_HGRE, GLRLM_SRHGE, GLRLM_RP, Race, Age, Tar_dose_min, Hypo_dose_median |
| FLAIR | min, std, max, Skewness, BinSize, Race, Growth Hormone, Age, Shunt, Tar_dose_min, Hypo_dose_min |

The robustness of ANN was tested with noise SNR of 20, 50, and 100 added to the radiomics features. As shown in Table 2, with high magnitude of noise with SNR 20, the accuracy decreased by 0.05, while for SNR 50 and 100, similar accuracy was achieved as ANN without noise. The impact of noise on key feature selection was shown in Fig. 4(D-F): the sparse distribution of the relative weights for each key feature were decreased by half the magnitude: dropping from 2.5 to 1.2, and the key features varied from original collection. Confusion matrices as shown in Fig. 5(D-F) suggest a high robust classification results for group 1 and 2, with consistent rates around 0.95, while the prediction accuracy for group 0 decreased by approximately 0.1 with the additional radiomic noise. The impact of radiomic noise on key feature selection and confusion matrix for T2w and FLAIR was similar to T1w as shown in Fig. 4 and 5, and the illustration could be found in Appendix FA1 and FA2.

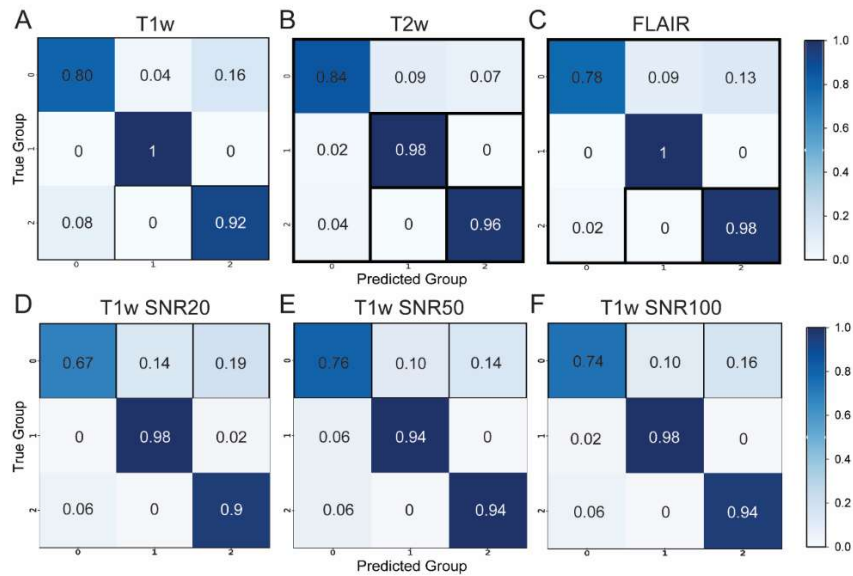

Fig. 5: Average confusion matrix over the 5-fold cross-validation (A-C) for T1w, T2w, and FLAIR MRI, and the noise effect (D-F) with SNR 20, 50, and 100 on T1w MRI.

**Discussion**

We defined the post-treatment obesity for pediatric craniopharyngioma patients as the swiped area between initial and follow-up BMI z-score as shown in Fig. 1(B). CDC [17] recommends 95[th] percentile or greater as the level above which children might have risk developing obesity-associated diseases. However, no clear guidelines for increment from baseline have been established. As a reference to the American Academy of Child and Adolescent Psychiatry, a child is generally not considered obese until their weight is at least 10% higher than that recommended for their height and body type. We selected 10 and 20 percentile thresholds to indicate the corresponding extent of BMI increase. This definition avoided a constant threshold that would distinguish two similar points across the threshold value, and reflected the extent of BMI increase after treatment. Thus, the endpoints were classified on normalized BMI increase instead of a thresholding BMI z-score.

Conventional statistical models including random forest or SelectKBest have been widely used in outcome predictions in radiation therapy. However, the prediction accuracy of these models is often sub-optimal and the key features selection has intrinsic limitation: each feature is sequentially added to or subtracted from the features pool and the optimal feature set is selected with the highest accuracy. This selection process only considers the impact of individual feature sequentially while the combination of certain features cannot be determined eg., feature A and B might be rejected as individual feature, while the combined set might boost the accuracy significantly. We applied ANN to overcome these limitations: the accuracy of ANN outperformed RF with clinical and dosimetric features by approximately 0.15, and RF with additional radiomic features by approximately 0.05; and the key features were selected with interplay, with clinical features consistent with our practical observation. However, the additional radiomic features increased the dimension of feature space and thus overfitting as a result. We used drop-out [30] regularization in each layer to mitigate the overfitting problem and achieved a relative stable classification accuracy during the 5-fold cross validation with a standard deviation of around 0.05. Moreover, since the Group lasso involved L1 norm indicated by the summation in Equation 2, the feature selection appeared more sensitive to noise when compared to the accuracy: even with a relatively small amount of noise of SNR 100, the key features would vary from original collection.

Clinical features including age, sex, race, shunt placement, and growth hormone therapy were highly correlated with post-treatment obesity in our clinic while there lacked quantitative statistical models. The ANN recognized these clinical key features among the high dimensional radiomic features, which justified the feature selection by Group lasso regularization. Only several dosimetric features were selected as key features mainly due to the uniform dose prescription in the patient cohort. Hypothalamus minimal dose was commonly selected as a key feature for all MRI, indicating the correlation between radiation caused hypothalamic damage and post-treatment obesity. The minimal dose to target was the dosimetric feature related to the re-occurrence of tumor and thus identified as key feature in FLAIR MRI. More aggressive dose constraint on hypothalamus is a potential mitigating method in treatment planning. Other common radiomic features including minimum gray level, histogram kurtosis and skewness could serve as imaging markers for post-treatment obesity prediction.

Although the multi-centric diagnostic MRI acquired for our patient cohort had intrinsic randomness, the limited data size and lack of external validation compromised the generalization of this study. Five-fold cross-validation randomly divided the 84 patients into the training and testing datasets in a ratio of 4:1, and every sample was rotating in the testing set. The standard deviation of accuracy in the cross-validation was around 4-7%, and resistant to noise. The average accuracy was compromised by 5% with noise of SNR20, while remained stable with SNR50 and SNR100. Future studies involving an expanded data size, and correlation of radiomic features with pathologic findings would benefit the early detection of post treatment side effect. The integration of radiomic information into treatment planning would direct the current physical dose-based to biological effect guided treatment planning system. The Group lasso regularization selected key features effectively while the noise resistance could be improved by replacing the L1 norm by higher order of summation to smooth the gradient in back-propagation.

**Conclusion**

Pediatric patients with craniopharyngioma treated with limited surgery and proton therapy often suffer from post-treatment obesity which compromises the quality of long-expected life span. However, the mechanism resulting such obesity or the intrinsic resistance to radiation therapy remains unclear. We developed a densely connected neural network to predict the probability of obesity, and identify key radiomic, clinical, and dosimetric features. The relative high prediction accuracy would indicate the necessity of interventional treatment, while the key features would indicate the methodology including conservative hypothalamus dose, initiation of growth hormone therapy, etc.